\documentclass{IEEEtran}
\usepackage{comment} 

\usepackage{cite}

\usepackage{amssymb,amsfonts}
\usepackage{textcomp}
\usepackage[pdftex]{graphicx}
\usepackage{algorithmic}
\usepackage{algorithm}
\usepackage{multicol}
\usepackage{bm}
\usepackage{lipsum, color}
\usepackage{commath}
\usepackage{epstopdf}
\usepackage{subfigure}
\usepackage{mathtools, nccmath}

\DeclareMathOperator*{\argmin}{argmin}
\begin{document}
\title{Hybrid Precoder and Combiner for Imperfect Beam Alignment in mmWave MIMO Systems}
\author{Chandan~Pradhan,~\IEEEmembership{Student Member,~IEEE,}
       Ang~Li,~\IEEEmembership{Member,~IEEE,}
       Li~Zhuo,~\IEEEmembership{Member,~IEEE,}
       Yonghui~Li,~\IEEEmembership{Fellow,~IEEE,}
       and~Branka~Vucetic,~\IEEEmembership{Fellow,~IEEE \vspace{-5.10ex}}
\thanks{The authors are  with  the  Centre  of  Excellence  in  Telecommunications, School of Electrical and Information Engineering, University of Sydney, Sydney, NSW 2006, Australia. (e-mail: \{chandan.pradhan,  ang.li2, yonghui.li, branka.vucetic\}@sydney.edu.au). Li Zhuo is with Beijing University of Technology, Beijing, China (email: zhuoli@bjut.edu.cn).}}

\maketitle

\begin{abstract}

In this letter, we aim to design a robust hybrid precoder and combiner against beam misalignment in millimeter-wave (mmWave) communication systems. We consider the inclusion of the `error statistics' into the precoder and combiner design, where the array response that incorporates the distribution of the misalignment error is first derived. An iterative algorithm is then proposed to design the robust hybrid precoder and combiner to maximize the array gain in the presence of beam misalignment. To further enhance the spectral efficiency, a second-stage digital precoder and combiner are included to mitigate the inter-stream interference. Numerical results show that the proposed robust hybrid precoder and combiner design can effectively alleviate the performance degradation incurred by beam misalignment.

\end{abstract}


\vspace{-4mm}

\section{Introduction}

\IEEEPARstart{M}ILLIMETER-WAVE (mmWave) communication has been identified as a potential technology capable of dispensing a large stretch of the underutilized spectrum ranging from 30 GHz to 300 GHz \cite{rappaport2013millimeter}. While the small wavelength of mmWave makes it vulnerable to path loss, penetration loss and rain fading \cite{rappaport2013millimeter}, it also allows the deployment of a large-scale antenna array  to exploit the high array gains to combat the  severe signal propagation loss in mmWave communications \cite{ hur2013millimeter, yu2016alternating}.

Unlike sub-6GHz commuications, the prohibitive cost and power consumption of the hardware components working at mmWave bands make hybrid processing a viable solution to reduce the number of RF chains at the transceivers, by performing signal processing in a low-dimensional digital domain and a high-dimensional analog domain \cite{el2014spatially, yu2016alternating, sohrabi2016hybrid, low2018Wang, beamSteering2012}.  Recent works on hybrid designs aim to maximize the overall spectral efficiency of the network with the assumption of perfect channel state information (CSI), which implicitly assumes perfect alignment between the transmitting and receiving beams. However, in practical mmWave scenarios where perfect CSI is usually not available \cite{el2014spatially}, the estimation errors in the angle of arrival (AoA) or angle of departure (AoD) result in beam misalignment. Additionally, the imperfection in the antenna array, which includes array perturbation and mutual coupling \cite{yang2016analysis}, and environmental vibrations such as wind, moving vehicles, etc., further contributes towards the imperfect alignment of beams \cite{hur2013millimeter}. Moreover, the deployment of a large-scale antenna array that generates narrow beams for mmWave communications also makes the system  highly sensitive to beam misalignment. This leads to a considerable loss in the array gain and consequently affects the system performance \cite{cheng2018coverage}. 

While there are already works that investigate the performance loss owing to beam misalignment \cite{yang2016analysis, cheng2018coverage}, there are only a limited number of studies that consider the robust hybrid designs in the presence of beam misalignment \cite{hur2013millimeter,Robust2018Precoding}. Moreover, these works  primarily focus on the single-receiver single-stream case only, and their extension to multi-stream communications is not straightforward. The statistics of the AoD/AoA estimation error have been studied in \cite{hur2013millimeter, Robust2018Precoding} and \cite{yu2006performance} and it is shown in \cite{Robust2018Precoding} that the inclusion of the `error statistics' into the hybrid design can lead to an improved performance in the case of beam misalignment. However, this concept has not been well explored for robust multi-stream hybrid precoder and combiner design in mmWave communications.

Motivated by this, in this paper we propose an iterative algorithm to alternatively design the robust hybrid analog-digital precoder and combiner against beam misalignment for a single-receiver multi-stream mmWave communication system. By incorporating the beam alignment error distribution, we first utilize the prior knowledge of the `error statistics' in beam misalignment to derive the expected array response for the transmitter and receiver, which form the basis for the row and columns space of the expected channel in the presence of beam misalignment \cite{beamSteering2012, el2014spatially}, respectively. Subsequently, we formulate the span for the feasible analog precoder and combiner with the digital precoder and combiner, respectively, which is projected onto the expected array response to maximize the array gain. The resulting optimization for the analog precoder and combiner is solved using the gradient projection (GP) method. To mitigate the inter-stream interference, a second-stage digital precoder and combiner is further included based on the obtained effective baseband channel. Numerical results show the desirable performance gains for the proposed robust design in the presence of imperfect beam alignment.  

\vspace{-2mm}

\section{System Model}

We consider a single-receiver mmWave system as shown in Fig.1, in which a base station (BS) with $M_t$ antennas transmits $N_s$ data streams to a receiver unit (RU) with $M_r$ antennas. The number of RF chains at the BS and RU is denoted by $N_{RF}^t$ and $N_{RF}^r$, respectively, where $N_s \leq N_{RF}^t \leq M_t$ and $N_s \leq N_{RF}^r \leq  M_r$.  During transmission, the BS employs a $N^t_{RF} \times N_s$ digital precoder $\mathbf{F_{BB}}=\left[{\bf f}_1^{\bf BB}, {\bf f}_2^{\bf BB}, \dots,{\bf f}_{N_s}^{\bf BB} \right]$  followed by an $M_t \times N^t_{RF}$ analog precoder $\mathbf{F_{RF}} = \left[{\bf f}_1^{\bf RF}, {\bf f}_2^{\bf RF}, \dots, {\bf f}_{N_{RF}^t}^{\bf RF}\right]$, and the transmitted signal can be written as $\mathbf{x} = \mathbf{F_{RF}F_{BB}s} = \mathbf{Fs}$, where $\mathbf{s}$ is the $N_s \times 1$ symbol vector and  $\mathbb{E}\left[{\bf s} {\bf s}^H \right] = \frac{P}{N_s} {\bf I}_{N_s}$. $P$ is the total transmit power at the BS, and in this work we have assumed uniform power allocation among different streams. Assuming the use of phase shifters for analog components, each entry of $\mathbf{F_{RF}}$ satisfies the element-wise constant-modulus constraint, i.e.,  $\left| \left[\mathbf{F_{RF}}\right]_{m,n} \right| = \sqrt{\frac{1}{M_t}}, \; \forall m,n$ \cite{ sohrabi2016hybrid}. The total power constraint is enforced by normalizing $\mathbf{F_{BB}}$ such that $\norm{\mathbf{F_{RF} F_{BB}}}^2_F = N_s$.  Considering a narrowband block fading propagation channel ${\bf H}$, the processed signal at the RU is given by   

\begin{equation}
\begin{split}
{\bf y} &= {\bf W}_{\bf BB}^H {\bf W}_{\bf RF}^{H} {\bf H} {\bf F}_{\bf RF} {\bf F}_{\bf BB} {\bf s} + {\bf W}_{\bf BB}^H {\bf W}_{\bf RF}^H {\bf n}, \\
\end{split}
\vspace{-4mm}
\end{equation}
where ${\bf W}_{\bf RF}^H$ is the $M_r \times N_{RF}^r$ analog combiner matrix with element-wise constant-modulus entries, i.e. $\left| \left[\mathbf{W_{RF}}\right]_{m,n} \right| = \sqrt{\frac{1}{M_r}}, \; \forall m,n$, ${\bf W_{BB}}$ is the low-dimensional digital combiner, and ${\bf n}$ is the noise vector with each entry following i.i.d $\mathcal{CN} (0, \sigma_n^2)$.
 
\vspace{-2mm}

\subsection{MmWave Channel Model}

MmWave channels are expected to be sparse with a limited number of propagation paths, given by \cite{rappaport2013millimeter}:

\begin{equation}
  \mathbf{H} = \sqrt{\frac{M_t M_r}{L}} \sum_{l = 1}^{L} {\gamma}_{l} {\bm \alpha} \left( \theta_l^{\left(AoA \right)} \right) {\bm \alpha} \left(\theta_l^{\left(AoD \right)} \right)^H, 
\end{equation}
where $L$ is the number of propagation paths between the BS and the RU, and ${\gamma}_{l}$ is the complex gain of the path following $\mathcal{CN}\left(0,\sigma^2_{\gamma}\right)$.  $\theta_l^{(AoD)}$ and $\theta_l^{(AoA)} \in [0, \pi]$ are the AoD and AoA, respectively, with ${\bm \alpha}\left(\theta_l^{(AoD)}\right)$ and ${\bm \alpha}\left(\theta_l^{(AoA)} \right)$ being the corresponding antenna array response vectors of the BS and RU, respectively. For uniform linear arrays (ULAs) considered in this paper, ${\bm \alpha}\left(\theta \right)$ for an $M$-element antenna array is given by

\begin{equation}
 {\bm \alpha}(\theta) = \frac{1}{\sqrt{M}} \left[1, e^{j \frac{2 \pi}{\lambda} d \; \cos(\theta)}, \dots, e^{j \frac{2 \pi}{\lambda} d \; (M - 1)\; \cos(\theta)} \right]^T,
\end{equation}
where $d$ and $\lambda$ are the antenna spacing and signal wavelength, respectively. 

\vspace{-4mm}

\subsection{Error Model for Beam Misalignment:} We define the beam misalignment error in AoA/AoD as $\delta \vcentcolon= \hat{\theta} -  \theta$, where $\hat{\theta}$ is the estimated AoA/AoD and $\theta$ is the actual AoA/AoD. The beam alignment error $\delta$ is characterized by a random variable following a uniform distribution as in \cite{Robust2018Precoding}, given by
\vspace{-4mm}

\begin{equation}
f(\delta) = 
\begin{cases}
    \frac{1}{2\beta},& \text{if } -\beta \leq \delta \leq \beta \\
    0,              & \text{otherwise,}
\end{cases}
\end{equation}
where $\beta = \sqrt{3} \Delta$ and $\Delta$ represents the standard deviation of the beam alignment error. The random misalignment error $\delta$ is bounded within the range of the mainlobe beamwidth $\vartheta$ of the transceiver units, i.e. $0 \leq |\delta| \leq \frac{\vartheta}{2}$, which is based on the fact that beam deviation exceeding the mainlobe beamwidth is treated as alignment failure rather than  misalignment \cite{yang2016analysis}.

\begin{figure}[t]
       \centering
        \includegraphics[width=3.6in, height=1.6in]{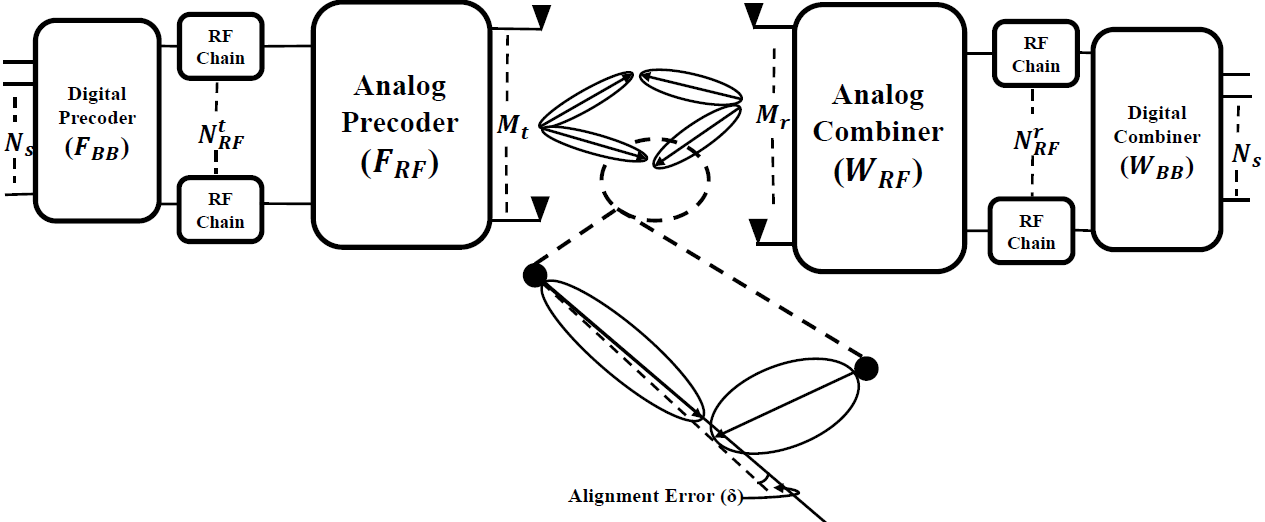}
    \caption{{ \small Block diagram for a point-to-point mmWave MIMO system}}
    \label{BD}
    \vspace{-6mm}
\end{figure}

\vspace{-4mm}

\section{Hybrid Precoder and Combiner Design}

To incorporate the effect of the beam misalignment, we first derive the  expected array response in the presence of beam alignment error before designing the robust precoder and combiner. Based on (3), the expected array response in the presence of beam alignment error $\delta$ is given by 
\begin{equation}
{{\bm \alpha^e}\left( \hat{\theta} \right)} = \sqrt{\frac{1}{M_t}} \left[1, \alpha^e_1(\hat{\theta)}, ...,  \alpha^e_{M_t}(\hat{\theta}) \right]^T,
\end{equation}
where $\alpha^e_m(\hat{\theta}) = \mathbb{E}\left[ e^{j \frac{2 \pi}{\lambda} d \; (m - 1) \cos(\theta + \delta)} \right], \; \forall m$. By defining $a_m \vcentcolon  = j \pi (m-1) \cos \theta$ and $b_m \vcentcolon  = j \pi  (m-1) \sin \theta$, $\alpha^e_m(\hat{\theta})$ is computed as

\begin{equation}
\begin{split}
\alpha^e_m(\hat{\theta}) &= \int_{-\beta}^{\beta} e^{j \pi (m-1) \cos(\theta + \delta)} f(\delta) d\delta \\
&= \frac{1}{2 \beta} \int_{-\beta}^{\beta} e^{a_m \cos \delta - b_m \sin \delta} d\delta \\
&\overset{(a)}\approx \frac{1}{2 \beta} \int_{-\beta}^{\beta} e^{a_m  - b_m \delta} d\delta \\
&\overset{(b)}= \frac{e^{a_m}}{\pi (m-1) \beta \sin \theta} \left( \frac{e^{j \pi (m-1) \beta \sin \theta} - e^{- j \pi (m-1) \beta \sin \theta}  } {2 j} \right) \\
&\overset{(c)}= e^{a_m} {\rm sinc} \left( (j \pi)^{-1} b_m \beta \right),
\end{split}
\end{equation}
where step (a) is inflicted from the small value of $\beta$, (b) is obtained from the definition of $b_m$ and (c) follows the Euler's formula with ${\rm sinc}(x) = \frac{\sin \pi x}{\pi x}$. Hereafter, we focus on the design for the robust hybrid precoder. The hybrid combiner design has a similar mathematical formulation and is therefore not included for the sake of  brevity. 

\vspace{-4mm}

\subsection{Hybrid Analog-Digital Precoder Design}

As observed from (5) and (6), ${\bm \alpha}^e\left( \hat{\theta}_l^{(AoD)}\right), \; \forall l$ in the presence of beam misalignment are no longer element-wise constant-magnitude, and hence cannot be directly used as the analog precoder as in \cite{beamSteering2012, el2014spatially}. Accordingly, we first find $N_{RF}^t$ dominant bases for the expected array response along $\hat{\theta}_l^{(AoD)}, \; \forall l$, followed by formulating the span for the analog precoder with the digital precoder. Subsequently, we employ a projection onto the dominant bases to maximize the array gain at the BS, while satisfying the element-wise constant-modulus constraint for the analog  precoder.  The hybrid precoder design is detailed below.

\subsubsection{Computation of $N_{RF}^t$ dominant basis for the expected array response} Defining the matrix ${\bf A}$ as 

\vspace{-2mm}

\begin{equation}
    {\bf A} = {\bm \Gamma} {\bf A}^t,
\end{equation}
where ${\bf A}^t = \left[ {\bm \alpha}^e\left(\hat{\theta}_1^{(AoD)}\right), {\bm \alpha}^e\left(\hat{\theta}_2^{(AoD)}\right), \dots, {\bm \alpha}^e\left(\hat{\theta}_L^{(AoD)}\right) \right]^H$ and ${\bm \Gamma} = diag \left[\gamma_1^*, \gamma_2^*, \dots, \gamma_L^* \right]$, we select the $M_t \times N_{RF}$ basis matrix ${\bf F}$ as the first  $N_{RF}^t$ column of ${\bf V}$, which is the right unitary matrix obtained from the singular value decomposition (SVD) of ${\bf A}$.  

\subsubsection{Projection of feasible hybrid precoder onto  ${\bf F}$ }

The span for the analog precoder ${\bf F}_{\bf RF}$ using the $N_{RF}^t \times N_{RF}^t$ digital precoder ${\bf \tilde{F}}_{\bf BB}$, i.e., ${\bf F}_{\bf RF} {\bf \tilde{F}}_{\bf BB}$ is projected onto ${\bf F}$  with ${\bf F}_{\bf RF}$ satisfying the element-wise constant-modulus constraint. Accordingly, we have the following optimization problem:

\vspace{-2mm}

\begin{equation}
\begin{aligned}
& \mathcal{P}_1: && \underset{{\bm F_ {\bm RF}}, {\bm \tilde{\bm F}}_ {\bm BB}}{\text{min}} \; \norm{ {\bf F}_{\bf RF} {\bf \tilde{F}}_{\bf BB} - {\bf F}}^2_F \\
& \text{\it s.t.}
& & {\rm C1}: \left| \left[\mathbf{F_{RF}} \right]_{m,n} \right| = \sqrt{\frac{1}{M_t}}, \; \forall m,n. 
\end{aligned}
\end{equation}
Basically, $\mathcal{P}_1$ is a  matrix  factorization  problem and is solved by alternately optimizing $\mathbf{F_{RF}}$ and $\mathbf{\tilde{F}_{BB}}$ \cite{yu2016alternating, el2014spatially}. To be more specific, the digital precoder $\mathbf{ \tilde{F}_{BB}}$ is designed based on a fixed analog precoder $\mathbf{F_{RF}}$ as an unconstrained least-square problem $\mathcal{P}_1$,  which leads to $\mathbf{\tilde{F}_{BB}} = {\bf F}_{\bf RF}^{\dagger} {\bf F}$. Subsequently, the analog precoder $\mathbf{F_{RF}}$ for a given $\mathbf{\tilde{F}_{BB}}$ is designed by solving the following sub-problem:

\vspace{-2mm}

\begin{equation}
\begin{aligned}
& \mathcal{P}_2: && \underset{{\bm F_ {\bm RF}}} {\text{min}} \; \norm{ {\bf F}_{\bf RF} {\bf \tilde{F}}_{\bf BB} - {\bf F}}^2_F \text{\it s.t.} \; {\rm C1}.
\vspace{-2mm}
\end{aligned}
\end{equation}
By defining $\mathbf{f} = {\rm vec}\big( {\bf F}\big)$, ${ {\bf B} = {\bf \tilde{F}}_{\bf BB}^T \otimes {\bf I}_{M_t}}$ and  $\mathbf{x} = {\rm vec}\big({\bf F_{RF}} \big)$, $\mathcal{P}_2$ is reformulated as the following constant-modulus least-square (CMLS) problem:

\begin{equation}
\begin{aligned}
& \mathcal{P}_{3}: && \underset{\text{\bf x}}{\text{min}} \; \norm{{{\bf B} {\bf x} -  {\bf f}}}^2_2 \\
& \text{\it s.t.}
& & {\rm C2}: \big|[{\bf x}]_{n}\big| = \sqrt{\frac{1}{M_t}}, \; n = 1,2,...M_t N_{RF}^t. 
\end{aligned}
\end{equation}
 In this work, $\mathcal{P}_{3}$ is solved using the GP method summarized in Algorithm \ref{GP_AG}. GP is a revamped version of conjugate gradient method \cite{rao2009engineering} which searches for the optimal solution in the decent direction by projecting each subsequent point onto the feasible region ${\rm C2}$, defined in Step 5, with the step size given by $\alpha_{GP} = \argmin_{\alpha \geq 0} \norm{{{\bf B} {\bf x} -  {\bf f}}}^2_2 \bigg|_{{{\bf x}^{(t)}} + \alpha {\bf d}^{(t)}}$. Finally, Algorithm 2 describes the framework to obtain the feasible hybrid  analog-digital precoder based on the principle of alternating optimization. 

\begin{algorithm}[!htb]
 \begin{algorithmic}[1]
\STATE \textbf{Input}: $\epsilon_{th}$, $ITR_{max}$
\STATE Initialize  $\epsilon^{(0)} \gets 0$, $\epsilon_d \gets 0$, $t \gets 0$, $\mathbf{x}^{(0)}$ with random phase, $\mathbf{d}^{(0)} \gets - 2{\bf B}^H \left({\bf B}{\bf x}^{(0)} - {\bf f} \right)$;
\WHILE {$\epsilon_d \geq \epsilon_{th}$ and $t \leq ITR_{max}$}
\STATE $\alpha_{GP} = \frac{\bigg(\mathcal{R} \left\{ {\bf f}^H {\bf B} {\bf d}^{(t)} \right\} - \mathcal{R} \left \{ \left({  {\bf d}^{(t)}} \right)^H {\bf B}^H {\bf B} {\bf x}^{(t)} \right\} \bigg)}{\left( \left({{\bf d}}^{(t)} \right)^H {\bf B}^H {\bf B} {\bf d}^{(t)} \right)}$;
\STATE ${\bf x}^{(t+1)} = \sqrt{\frac{1}{M_t}} e^{j \angle {\left( {\bf x}^{(t)} + \alpha_{GP} \mathbf{d}^{(t)} \right)}} $;
\STATE ${\nabla {\bf g} \left({\bf x}^{(t+1)} \right)} = 2{\bf B}^H \left({\bf B}{\bf x}^{(t+1)} - {\bf f} \right)$; 
\STATE $\beta_{GP} = \frac{ \left({\nabla {\bf g} \left({\bf x}^{(t+1)} \right)} - {\nabla {\bf g} \left({\bf x}^{(t)} \right)} \right)^H {\nabla {\bf g} \left({\bf x}^{(t+1)} \right)}}{ \left({\nabla {\bf g} \left({\bf x}^{(t)} \right)} \right)^H {\nabla {\bf g} \left({\bf x}^{(t)} \right)} }$;
\STATE ${\bf d}^{(t+1)} = - {\nabla {\bf g} \left({\bf x}^{(t+1)} \right)} + \beta_{GP} \mathbf{d}^{(t)}$;
\STATE $\epsilon^{(t+1)} = \norm{{{\bf B} {\bf x} -  {\bf f}}}^2_2$; 
\STATE $\epsilon_d = \left| \epsilon^{(t+1)} - \epsilon^{(t)} \right| $; $t \gets t + 1$.
\ENDWHILE
\STATE \textbf{Output}: ${{\bf x}^{(t+1)}}$
\end{algorithmic}
\caption{Analog Precoder Design - Gradient Projection}
\label{GP_AG}
\vspace{-0.35em}
\end{algorithm}

\begin{algorithm}[!htb]
 \begin{algorithmic}[1]
\STATE \textbf{Input}: ${\bf A}$, $\epsilon_{th}$, $ITR_{max}$
\STATE Initialize $\epsilon_d \gets 1$, $t \gets 1$, $\epsilon^{(0)} \gets 10$;
\STATE SVD decomposition: ${\bf A} = {\bf U} {\bf \Sigma} {\bf V}^H$;
\STATE Initialize ${\bf F} \gets {\bf V}(:,1:N_{RF}^t)$, ${\bf F}_{\bf RF}^{(0)} \gets \frac{1}{\sqrt{M}} e^{j \angle {\bf F} }$; 
\WHILE {$\epsilon_d \geq \epsilon_{th}$ and $t \leq ITR_{max}$}
\STATE ${{\bf \tilde{F}}^{(t)}_{\bf BB}} = \left( {{\bf F}^{(t-1)}_{\bf RF}} \right)^{\dagger} {\bf F}$;
\STATE Update ${{\bf F}^{(t+1)}_{\bf RF}}$ using GP;
\STATE $\epsilon^{(t+1)} = \norm{{ {{\bf F}^{(t+1)}_{\bf RF}} {{\bf \tilde{F}}^{(t+1)}_{\bf BB}} } - {\bf F}}_F^2$;
\STATE $\epsilon_d = \left|\epsilon^{(t)} - \epsilon^{(t-1)} \right|$, $t \gets t + 1$;
\ENDWHILE
\STATE \textbf{Output}: $\mathbf{F}_{\bf RF}, \mathbf{\tilde{F}}_{\bf BB}$
\end{algorithmic}
\caption{Alternate Analog-Digital Precoder Design}
\label{HADP_AG}
\vspace{-0.35em}
\end{algorithm}

\vspace{-4mm}

\subsection{Second-Stage Digital Precoder and Combiner Design}

We further introduce a second-stage digital precoder and combiner to cancel the inter-stream interference. For a given hybrid analog-digital precoder and combiner, we can obtain the effective baseband channel ${\bf H}_{\bf e}$ as \cite{low2018Wang}

\begin{equation}
    {\bf H}_{\bf e} = {\bf \tilde{W}}_{\bf BB} {{\bf W}_{\bf RF}}  {\bf H}  {{\bf F}_{\bf RF}} {\bf \tilde{F}}_{\bf BB},
\end{equation}
and we define the SVD of the effective baseband channel ${\bf H}_{\bf e}$ as ${\bf H}_{\bf e} = {\bf U}_{\bf e} {\bm \Sigma}_{\bf e} {\bf V}^H_{\bf e}$. Then an SVD-based second-stage digital precoder and combiner are obtained as

\begin{equation}
    {\bf \bar{F}}_{\bf BB} = {\bf V_e}(:, 1:N_s), {\bf \bar{W}}_{\bf BB} = {\bf U_e}(:, 1:N_s),
\end{equation}
and the effective robust digital precoder and combiner are given by \cite{low2018Wang}
\vspace{-2mm}


\begin{equation}
    {\bf F}_{\bf BB} = \sqrt{N_s}  \frac{{\bf \tilde{F}}_{\bf BB} {\bf \bar{F}}_{\bf BB}} {\norm{{\bf F}_{\bf RF} {\bf \tilde{F}}_{\bf BB} {\bf \bar{F}}_{\bf BB}}_F^2},  {\bf W}_{\bf BB} = {\bf \tilde{W}}_{\bf BB} {\bf \bar{W}}_{\bf BB}.
\end{equation}

 
The final iterative algorithm to design the robust analog-digital precoder and combiner is summarized in Algorithm 3.

\begin{algorithm}[h]
 \begin{algorithmic}[1]
\STATE \textbf{Input}:  $\hat{\theta}_l^{(AoD)}, \; \hat{\theta}_l^{(AoA)}, \; \forall l$,  $\epsilon_{th}$
\STATE Initialize $\epsilon_d \gets 1$, $t \gets 1$;
\STATE Initialize ${\bf \tilde{W}}_{\bf BB}^{(0)}$ and ${{\bf W}_{\bf RF}^{(0)}}$ using random phase;
\WHILE {$t \leq 2$}
\STATE Obtain ${{\bf F}_{\bf RF}^{(t)}}$ and ${\bf \tilde{F}}_{\bf BB}^{(t)}$ using Algorithm 2;
\STATE Calculate ${\bf H}_{\bf e} = {\bf \tilde{W}}_{\bf BB}^{(t-1)} {{\bf W}_{\bf RF}^{(t-1)}}  {\bf H}  {{\bf F}_{\bf RF}^{(t)}} {\bf \tilde{F}}_{\bf BB}^{(t)}$;
\STATE ${\bf \bar{F}}_{\bf BB}^{(t)} = {\bf V}_{\bf e}(:, 1:N_s), \; {\text s.t.} \; {\bf H}_{\bf e} = {{\bf U}_{\bf e} {\bf \Sigma}_{\bf e} {\bf V}^H_{\bf e}}$;
\STATE ${\bf F}_{\bf BB}^{(t)} = \sqrt{N_s}  \frac{{\bf \tilde{F}}_{\bf BB}^{(t)} {\bf \bar{F}}_{\bf BB}^{(t)}} {\norm{{\bf F}_{\bf RF}^{(t)} {\bf \tilde{F}}_{\bf BB}^{(t)} {\bf \bar{F}}_{\bf BB}^{(t)}}_F^2}$;
\STATE Obtain ${{\bf W}_{\bf RF}^{(t)}}$ and ${\bf \tilde{W}}_{\bf BB}^{(t)}$ using Algorithm 2;
\STATE Calculate ${\bf H}_{\bf e} = {\bf \tilde{W}}_{\bf BB}^{(t)} {{\bf W}_{\bf RF}^{(t)}}  {\bf H}  {{\bf F}_{\bf RF}^{(t)}} {\bf \tilde{F}}_{\bf BB}^{(t)}$;
\STATE ${\bf \bar{W}}_{\bf BB}^{(t)} = {\bf U}_{\bf e}(:, 1:N_s), \; {\text s.t.} \;  {\bf H}_{\bf e} = {{\bf U}_{\bf e} {\bf \Sigma}_{\bf e} {\bf V}^H_{\bf e}}$;
\STATE ${\bf W}_{\bf BB}^{(t)} =  {\bf \tilde{W}}_{\bf BB}^{(t)} {\bf \bar{W}}_{\bf BB}^{(t)}$,  $t \gets t + 1$.
\ENDWHILE
\STATE \textbf{Output}: ${\bf F}_{\bf RF}$, ${\bf F}_{\bf BB}$, ${\bf W}_{\bf RF}$, ${\bf W}_{\bf BB}$
\end{algorithmic}
\caption{Robust Hybrid Precoder-Combiner Design}
\label{HJHP_AG}
\vspace{-0.35em}
\end{algorithm}

\vspace{-3mm}

\section{Numerical Results} 

In  this  section,  we  evaluate  the  performance of our proposed scheme via Monte-Carlo simulations. Unless stated otherwise, we assume $M_t = 128$, $M_r = 72$, $N_{s} = 4$,  $L = 10$ for the mmWave channel, and  the beam alignment error with a standard deviation of $\Delta = 1.154$, i.e.,   $-2^o \leq \delta \leq 2^o$. SNR is defined as $ {\rm SNR} = 10 \; \log_{10} \frac{1}{\sigma^2}$, where the total transmit power is set as $P = 1$. The  antenna spacing is $d = \frac{\lambda}{2}$  and  all  simulation  results  are  averaged  over  ${\rm 10^3}$ channel realizations. We use the following schemes as benchmarks: 1) Robust Fully-digital Precoder (R-DB) obtained from $ \mathbf{H}^e = \sqrt{\frac{M_t M_r}{L}} \sum_{l = 1}^{L} {\gamma}_{l} {\bm \alpha}^e \left( \hat{\theta}_l^{\left(AoA \right)} \right) {\bm \alpha}^e \left(\hat{\theta}_l^{\left(AoD \right)} \right)^H$, 2) Non-robust Fully-digital Precoder (NR-DB) obtained from ${\bf H}$, 3) Non-robust Hybrid Precoder (NR-HYB): Hybrid Precoder and combiner discussed in Section III without incorporating the error-statistics into the design, 4) SOMP: Hybrid precoder and combiner design proposed in \cite{el2014spatially}, and 5) PE-AltMin: Hybrid precoder and combiner design proposed in \cite{yu2016alternating}.

 The spectral efficiency v.s. $N_{RF}$ performance of the proposed hybrid design with $N^t_{RF} = N^r_{RF} = N_{RF}$ and at ${\rm SNR} = 0 {\rm \; dB}$ is shown in Fig.2. It is observed that the proposed robust hybrid precoder and combiner design, denoted by R-HYB, enjoys a noticeable gain over the non-robust hybrid designs for $N_{RF} \geq N_s$.

\vspace{-4mm}
\begin{figure}[!htb]
    \centering
        \includegraphics[scale = 0.24]{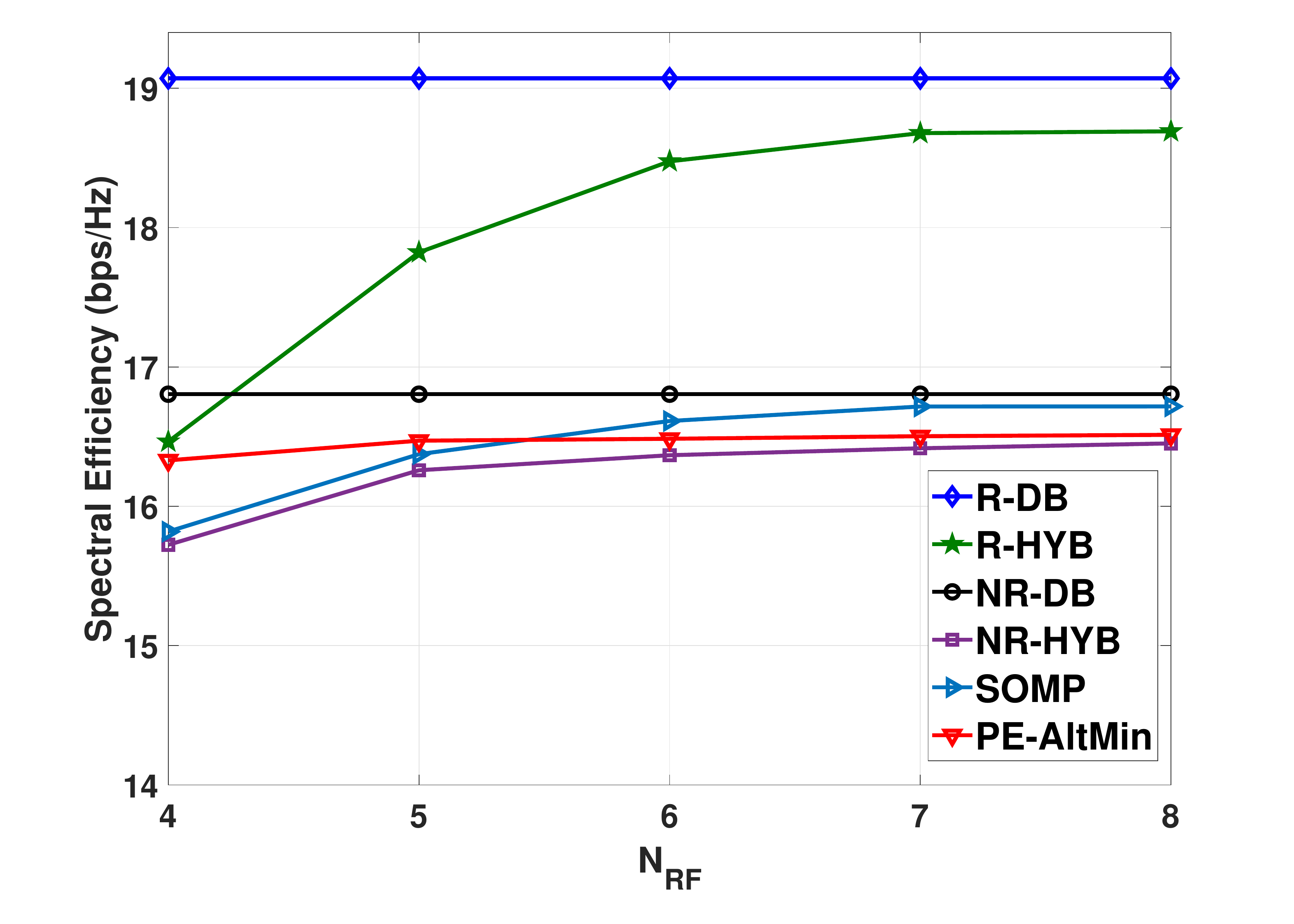}
        \vspace*{-3mm}
    \caption{\small Spectral efficiency v.s. $N_{RF}$, $M_t = 128$, $M_r = 72$, $N_s = 4$, ${\rm SNR} = 0 \; {\rm dB}$ and $N^t_{RF} = N^r_{RF} = N_{RF}$.}
    \label{BP}
    \vspace{-2mm}
\end{figure}

Fig.3 presents the spectral efficiency v.s. SNR performance of the proposed hybrid precoder and combiner with $N^t_{RF} = N^r_{RF} = 8$. At the high SNR, it can be observed that the proposed robust hybrid design achieves a gain of at least $2 \; {\rm dB}$ over the non-robust designs in the presence of the beam misalignment. 

\vspace{-4mm}

\begin{figure}[!htb]
    \centering
   
        \includegraphics[scale = 0.24]{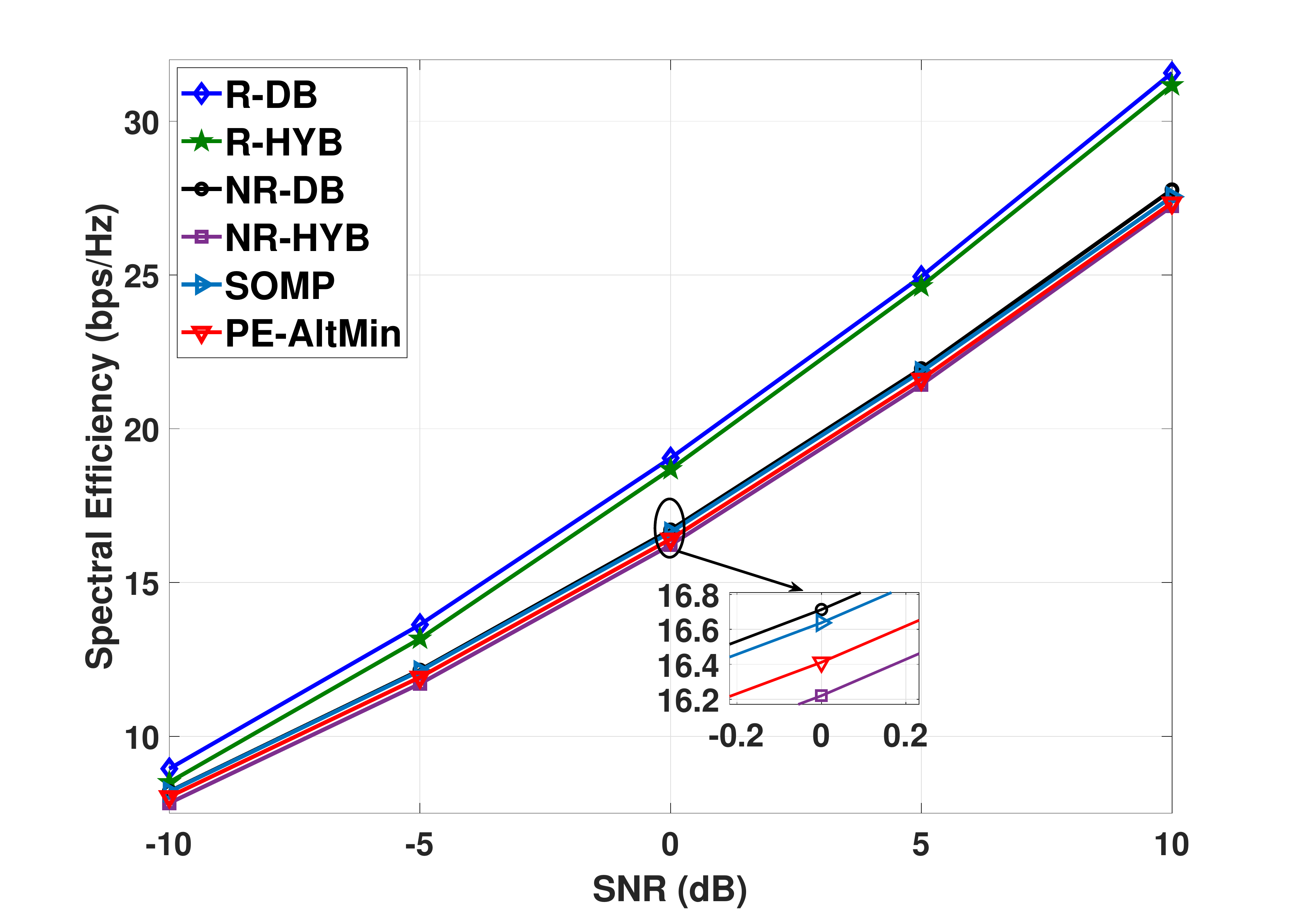}
        \vspace*{-3mm}
    \caption{\small Spectral Efficiency v.s. SNR, $M_t = 128$, $M_r = 72$, $N_s = 4$ and  $N^t_{RF} = N^r_{RF} = 8$.}
    \label{BP}
    \vspace{-2mm}
\end{figure}

\vspace{-0mm}

\section{Conclusion}

In this paper, we have proposed a robust hybrid precoder and combiner based on `error-statistics' to abate the performance loss owing to imperfect alignment between the beams at the transmitter and receiver. The  robustness of the proposed design has been validated through numerical examples and it has been shown to enjoy considerable performance gains compared to its non-robust counterparts. 

\vspace{-2mm}

\bibliography{main.bib}
\bibliographystyle{IEEEtran}

\end{document}